\documentclass{article}

\usepackage{arxiv}

\usepackage[utf8]{inputenc} 
\usepackage[T1]{fontenc}    
\usepackage{hyperref}       
\usepackage{url}            
\usepackage{booktabs}       
\usepackage{amsfonts}       
\usepackage{nicefrac}       
\usepackage{microtype}      
\usepackage{lipsum}
\usepackage{graphicx}
\usepackage{authblk}
\usepackage{gensymb}
\usepackage{float}
\graphicspath{ {./images/} }
\usepackage{mathtools}
\usepackage{amsmath,amssymb}
\usepackage{comment}
\usepackage{mathrsfs}
\usepackage{graphicx,nicefrac}
\usepackage{float}
\usepackage{amssymb}

\usepackage{comment}
\usepackage{scalerel}
\usepackage[overload]{empheq}
\usepackage{textgreek}

\usepackage{algorithm}
\usepackage{algpseudocode}
\usepackage{enumitem}

\usepackage{mathrsfs}
\usepackage{graphicx,nicefrac}

\usepackage[utf8]{inputenc}
\usepackage[english]{babel}

\renewcommand{\baselinestretch}{0.9}

\usepackage[margin=1cm]{caption}
\usepackage{comment}
\usepackage{scalerel}
\usepackage[overload]{empheq}

\usepackage{float}

\usepackage{mathtools}

\providecommand{\keywords}[1]
{
  \small	
  \textbf{\textit{Keywords---}} #1
}

\usepackage[nocompress]{cite}
\usepackage{url}
\usepackage{amsmath,amssymb,graphicx}

\usepackage{lipsum} 
\usepackage{mathrsfs,amsmath}
\usepackage{comment}
\usepackage{scalerel}
\usepackage[overload]{empheq}
\usepackage{nicefrac,xfrac}
\usepackage{mathtools}
\usepackage{gensymb}
\usepackage{soul}
\usepackage{textgreek}

\usepackage{amsmath}
\usepackage{bm}
\usepackage{algorithm}
\usepackage{algpseudocode}
\usepackage{enumitem}
\usepackage{amsmath}
\usepackage{setspace}
\usepackage{xcolor}
\usepackage{gensymb}
\usepackage{multicol,lipsum}

\usepackage{mathtools}
\usepackage{cuted}

\usepackage[switch]{lineno}

\usepackage{siunitx}
\DeclareSIUnit\pixel{pixel}
\DeclareSIUnit\inch{in}
\DeclareSIUnit\FPS{FPS}

\usepackage{wasysym}
\usepackage{soul}

\usepackage{authblk}
\title{Multi-modal imaging using a cascaded microscope design}

\author[1]{Xi Yang}
\author[2]{Mark Harfouche}
\author[1]{Kevin C. Zhou}
\author[1]{Lucas Kreiss}
\author[1]{Shiqi Xu}
\author[1]{Kanghyun Kim}
\author[1]{Roarke Horstmeyer}

\affil[1]{Duke University, Department of Biomedical Engineering}
\affil[2]{Ramona Optics, Inc.}

\affil[*]{Corresponding author: roarke.w.horstmeyer@duke.edu}

\begin{document}
\maketitle

\begin{abstract}
We present a new \emph{Multimodal Fiber Array Snapshot Technique (M-FAST)}, based on an array of 96 compact cameras placed behind a primary objective lens and a fiber bundle array. which is capable of large-area, high-resolution, multi-channel video acquisition. The proposed design provides two key improvements to prior cascaded imaging system approaches: a novel optical arrangement that accommodates the use of planar camera arrays, and the new ability to acquire multi-modal image data acquisition. M-FAST is a multi-modal, scalable imaging system that can acquire snapshot dual-channel fluorescence images as well as d phase contrast measurements over a large 8~x~10~mm$^2$~FOV at 2.2~$\mu$m full-pitch resolution. 
\end{abstract}


Large area, high-resolution microscope imaging plays an increasingly important role in basic research, clinical workflows, and industrial inspection. Large field-of-view (FOV) microscopes, capable of resolving more than hundreds of millions pixels per snapshot, are needed to efficiently image whole-mount tissue specimens \cite{sdobnov2018recent}, e.g., to monitor different neurons over an entire mouse cortex \cite{kim2016simultaneous,fan2019video}, and to observe freely moving organisms at cellular level detail \cite{thomson2021gigapixel,krishnamurthy2019scale,JOHNSON202070,nguyen2016whole}. Standard microscopes that were historically designed for the human visual system and based on single objective lenses often cannot meet the needs of such novel applications. Moreover, there is an increasing demand for acquisition of \emph{multi-modal} image data, including multi-channel fluorescence, phase contrast information, and polarimetric data, which is challenging to integrate into traditional systems on a large scale. 
It is well-known\cite{park2021review} that the number of resolved pixels per image, otherwise referred to as the system space bandwidth product (SBP), is practically limited by two components within a standard digital microscope: the microscope objective lens and the digital image sensor. Over the past few years, there have been several computational imaging-based approaches to increase an imaging system's SBP to simultaneously provide high resolution over a large FOV, including Fourier ptychography~\cite{konda2020fourier} and its vectorial variation~\cite{dai2022quantitative,xu2021imaging}, large-scale scanning light-sheet microscopes~\cite{voigt2019mesospim} and multi-view microscopes\cite{he2020snapshot}. While with high SBP, such approaches typically must acquire multiple snapshots over time and cannot resolve rapidly moving specimens or transient dynamics. 

For single-snapshot high-SBP acquisition, explored strategies include direct optimization of the objective lens itself \cite{mcconnell2016novel} and the use of multiple imaging systems, or "multi-scale" systems \cite{brady2012multiscale}, that are arranged in a cascaded fashion to acquire data in parallel. A recent work applied such parallelized "multiscale" concepts~\cite{park2021review} to create the RUSH system~\cite{fan2019video}, a video-rate large-SBP microscope with a custom-designed objective lens, which forms a magnified spherical intermediate image. A number of individual compact sCMOS cameras then re-image unique segments of this intermediate surface, and the resulting data is stitched into a final composite. While the use of a curved intermediate image plane in such prior designs avoids the introduction of multiple sources of aberration \cite{brady2012multiscale}, it also introduces several critical engineering challenges, including tilted cameras overlapping in a common focal plane design and tight camera packing density. As we will detail, these two aspects present significant limitations to the goal of multi-modal acquisition (e.g., the ability to capture multiple fluorescence emission wavelengths per snapshot).

\begin{figure}[!t]
\centering
\fbox{\includegraphics[width=7.8cm]{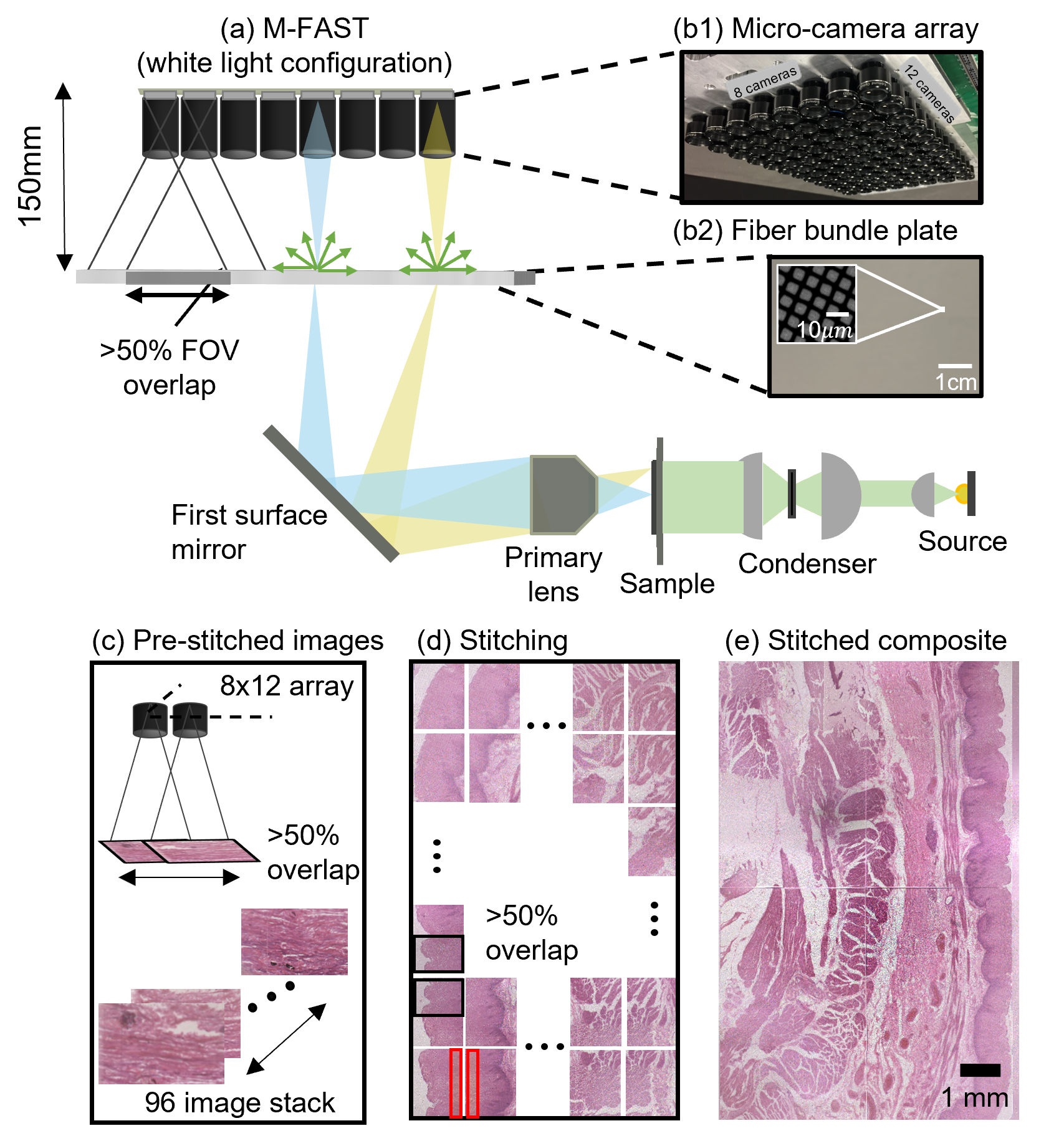}}
\caption{(a) Schematic of the M-FAST microscope. (b) Image of micro camera array (top) and of fiber bundle array plate (bottom). (c) Captured 96 ($8 \times 12$) raw images of a lung histology sample with $>$\SI{50}{\percent} FOV overlap in one dimension (black box), to be stitched in (d). (e) Large-FOV stitched composite, high resolution composite of thin-tissue specimen.}
\label{fig:setup}
\end{figure}

In this work, we build upon such prior designs to produce a scalable, cascaded microscope design that facilitates multi-modal image acquisition, including both phase contrast as well as dual-channel fluorescence imaging. Our new "Multimodal Fiber Array Snapshot Technique" (M-FAST) is composed of three main components: a primary objective lens, a flat integrated micro-camera array board \cite{thomson2021gigapixel} and a fiber bundle array plate. The \textbf{primary objective lens} (finite conjugate lens: RMA Electronics, FL-CC0814-5M, 0.36 numerical aperture~(NA) ) offers an optical SBP of several hundred megapixels~\cite{zheng20140} and magnifies the sample onto an intermediate plane ($8\times10\text{mm}^2$~FOV, $16\times24\text{cm}^2$ magnified image size). The intermediate plane contains the \textbf{fiber faceplate array (FFA)} (details below). Light emerging from the fiber array is then recorded by a \textbf{micro-camera array}. Each micro-camera is configured to focus on the intermediate image plane and capture data from a unique FOV that overlaps with those of neighboring cameras, as sketched in Fig.~\ref{fig:setup}(a). The cameras' magnification was designed to have an approximately \SI{10}{\percent} FOV overlap along the short side of the rectangular sensor (Fig.~\ref{fig:setup}(d) red box) and \SI{52}{\percent} overlap along the longer dimension (Fig.~\ref{fig:setup}(d) black box), ensuring each point within the intermediate image is observed at least by two micro-cameras for multi-channel acquisition.

\begin{figure}[!b]
\centering
\fbox{\includegraphics[width=7.8cm]{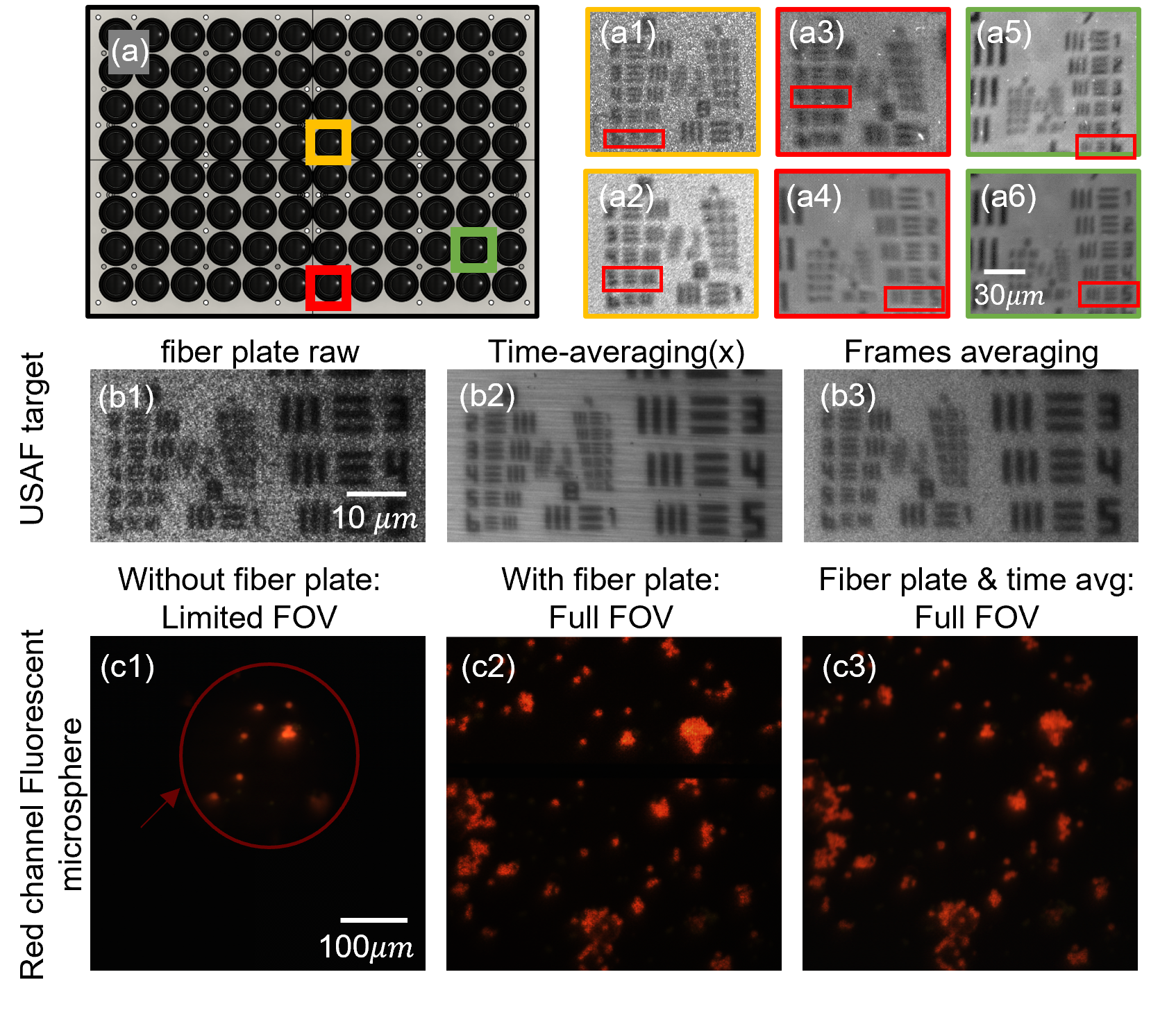}}
\caption{(a) Photo of $8\times12$ micro-camera array. (a1-a6) Bright-field images from both center (first row) and corner (second row) of three unique micro-camera images, each marked with colored box. (b1-b3) Bright-field images of resolution target with fiber faceplate raw, time-averaging, and frame averaging. (c1) and (c2) shows impressive vignetting alleviation with the FFA. Red circle in (c1) shows the vignetting effect while being eliminated in (c2). Translating the FFA \SI{1}{\milli\meter} within a \SI{0.2}{\second} exposure time leads to the example time-averaged result in Fig.~\ref{fig2:resolution}(c3).}
\label{fig2:resolution}
\end{figure}

\begin{figure*}[h!]
\centering\includegraphics[width=0.9\linewidth]{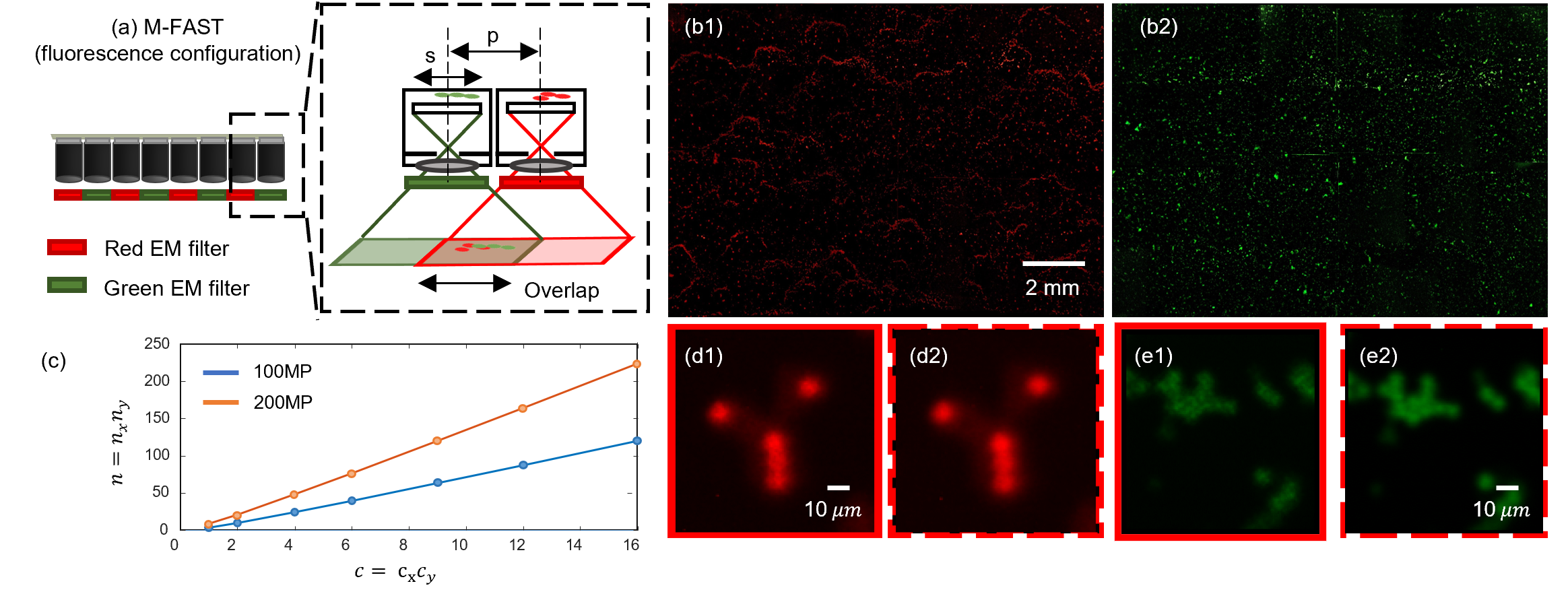}
\caption{(a) Dual channel fluorescence imaging set up schematic showing \SI{55}{\percent} overlap along $x$ and alternating red/green emission filters inserted over adjacent micro-cameras to concurrently detect red/green fluorescence over entire FOV. (b) Snapshot dual-channel fluorescence image acquired by the M-FAST ($8\times10\text{mm}^2$  FOV) of red (b1)/green (b2) fluorescent microspheres with zoomed version in (d)/(e). (d1)(e1) shows the raw results with speckle noise while (d2)(e2) shows the time-averaged results. (c) Plot of required number of micro-cameras as a function of desired number of multi-mode channels for standard M-FAST parameters assuming a primary lens with 11cm by 11cm FOV (blue) and 22cm by 22cm FOV megapixel (orange) SBP.}
\label{fig:fluourscent}
\end{figure*}

A key component of our new design is the FFA, which is a thin, large plate of fused glass single-mode fibers. The FFA accounts for the primary objective lens' non-telecentricity. Ideally, light cones at the intermediate image plane formed by an image-side telecentric primary lens would align with the optical axis of each micro-array camera, thus producing an image without vignetting. Unfortunately, it is significantly challenging and expensive to design microscope objective lenses with a high SBP and image-side telecentricity. To decouple multiscale microscopes from such a challenging requirement, we instead adopted a non-telecentric primary objective lens whose chief rays on the image side are not parallel (see sketch in Fig.~\ref{fig:setup}(a)). Recent works show sub-micrometer resolution imaging\cite{liu2021meta}, wide angle imaging\cite{stamenov2014panoramic} with a fiber bundle based system. In addition, it has been established that speckle sensing methods can be used to image dynamics~\cite{boas2010laser,xu2021imaging2,xu2022transient}. In a configuration without FFA, off-axis light from the primary lens is traveling at an angle that can exceed the NA of micro-cameras when the micro-cameras are located in the peripheral array. This leads to image signal reduced or vanished. By introducing a FFA at the intermediate image plane, light is re-directed to enter the lens of each single micro-camera of the array to form an image with minimal vignetting. Fig.~\ref{fig2:resolution}(c1)(c2) highlights the benefit of adopting the FFA. Without the array, vignetting severely limits the observable micro-camera FOV of this exemplary 10 $\mu$m fluorescent microsphere calibration specimen. In the current version of our M-FAST microscope, we use a \SI{6}{\micro\meter} size fiber faceplate (SCHOTT, glass type: 47A) with 5~mm thickness and per-fiber NA=1. We placed four such arrays, each $8~\times~12$~cm in size, directly against one another to cover the M-FAST intermediate image plane. Fig.~\ref{fig:setup}(b2) shows the \SI{6}{\micro\meter} fiber faceplate under a $40\times$ microscope.  

The micro-camera array contains 96 individual imagers ($8\times12$ grid, \SI{19}{\milli\meter} pitch), each consisting of a rectangular CMOS sensor (Omnivision 108238, 4320 x 2432 pixels with a pixel size of 1.4 um) that are jointly controlled by a custom electronics arrangement to simultaneously acquire image data from all sensors (\SI{0.96}{gigapixels} acquired per snapshot). Each micro-camera utilizes a \SI{25}{\milli\meter} focal length lens ($f/2.5$, Edmund Optics) configured to image at 0.14 magnification. Each micro-camera's NA is 0.032, which was selected to approximately match the image-side NA of the primary objective lens. Stitching the images captured by all 96 micro-cameras (via PTGui software \cite{PTGui}) produces a final full-FOV composite. Unlike prior multiscale microscope designs, the M-FAST micro-cameras are arranged to capture FOVs that exhibit a large degree of overlap with their immediate neighbors to facilitate snapshot multimodal acquisition. In our current design, we aimed to capture two modalities per snapshot: each point on the intermediate image plane is observed by at least two micro-cameras. Within this configuration, the half-pitch resolution of the micro camera array at the intermediate image plane was 9 $\mu$m, when directly imaging a target at the intermediate image plane without a primary lens. The complete M-FAST system has 3.81 magnification. The entire FOV of the complete system is 8x10~mm. To assess the resolution of our M-FAST microscope. We translated a USAF resolution target (Ready Optics) via a manual 3-axis stage to different FOV areas. Fig.~\ref{fig2:resolution} (a) shows bright-field imaging results of the resolution target within the FOV of three different micro-cameras with uniquely colored outlines. At the center of the FOV of the central micro-camera, we measured a \SI{2.2}{\micro\meter} two-point resolution, which drops to \SI{2.76}{\micro\meter} at the edge of the central micro-camera's FOV. We observed a \SI{3.1}{\micro\meter} and \SI{3.48}{\micro\meter} two-point resolution at the FOV center and edge, respectively, for the micro-camera at the array edge. 

There are two key issues that impact image fidelity within our unique optical arrangement that we address using software post-processing. First, vignetting effects introduced by the primary objective lens and camera array lenses lead to a fixed intensity fall-off towards the FOV edges of each camera as well as of the composite system. We calibrate the images based on a Gaussian-blurred blank reference image. The second issue is the introduction of fixed, high-spatial frequency modulation by the fiber faceplate, whose fibers (6 $\mu$m average diameter) are smaller than the M-FAST system's resolution at the intermediate image plane (9 $\mu$m half-pitch resolution), yet still lead to a speckle-like modulation effect, as observable in Fig.~\ref{fig2:resolution}(b1). To remove these modulation effects, We utilized a motorized stage to vibrate the fiber faceplate \cite{fujimaki2014reduction} during the finite image exposure. We adopted two methods to achieve fiber faceplate vibration: time averaging Fig.~\ref{fig2:resolution}(b2) and frame averaging Fig.~\ref{fig2:resolution}(b3). For the time averaging method, we vibrated the fiber faceplate within a fixed exposure time, while for frame averaging we captured and computed the average of $10$ frames, each with the plate at a random displacement (\SIrange{1}{3}{\milli\meter})(see Fig.~\ref{fig2:resolution}(b3)). In the future, calculating the transmission matrix of the fiber faceplate could enable effective removal without moving parts in the system \cite{huang2020retrieving}.


\begin{figure}[ht!]
\centering
\fbox{\includegraphics[width=7.8cm]{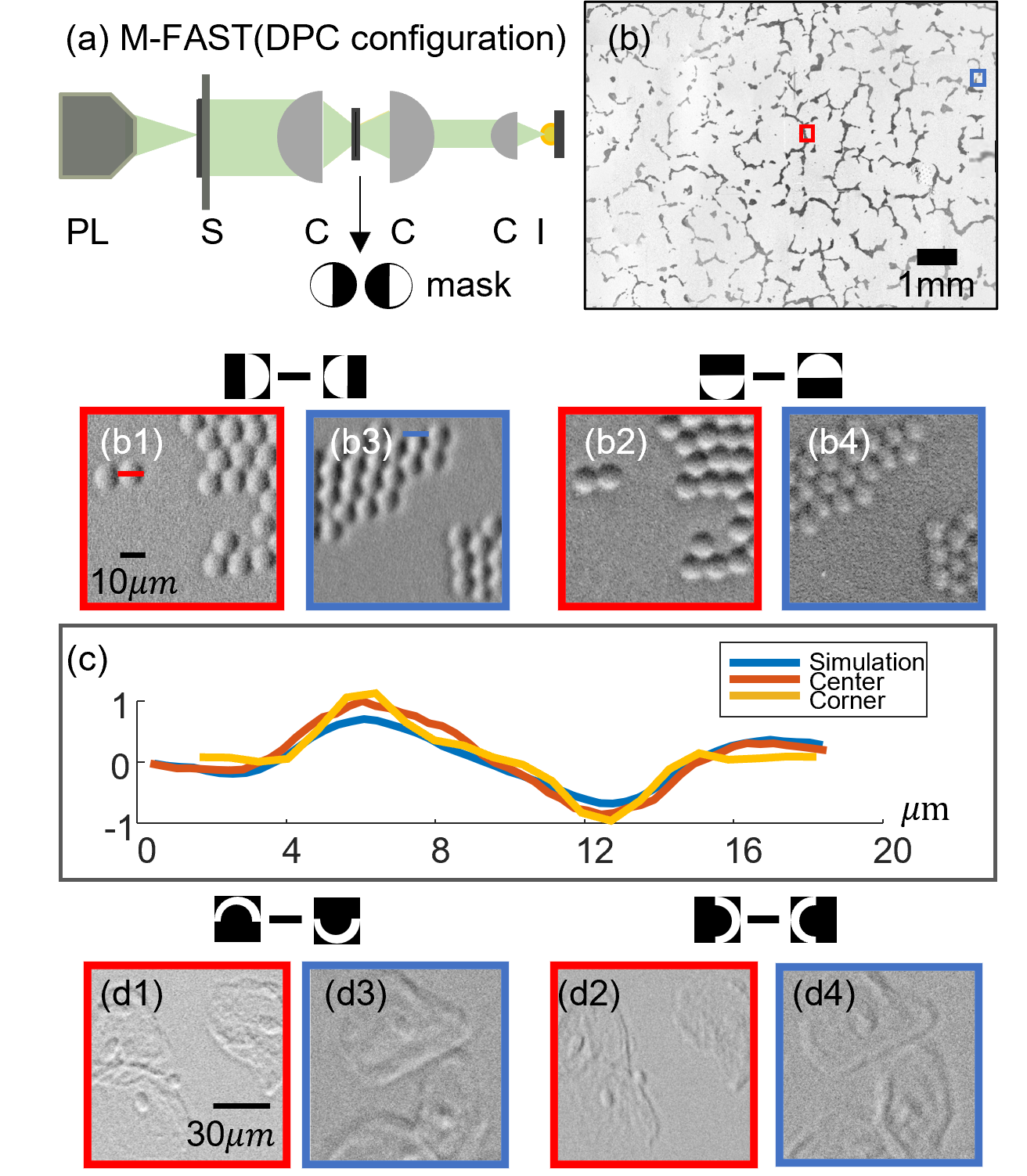}}
\caption{(a) M-FAST differential phase contrast configuration, with a illumination LED source(I), condensers(C), sample(S) and primary lens(PL). The FFA and micro-camera array are not shown here. (b) Bright field image of dense collection of 10$\mu$m microspheres immersed in n=1.56 oil. DPC results shown for zoomed-in areas at FOV center (b1)-(b2) and corner (b3)-(b4). (c) 1D surface profiles of microspheres within DPC image (averaged over 20 beads) at FOV corner (blue) and FOV center(red), along with the 1D simulated surface profile from known sphere thickness profile. (d) DPC image of Oral epithelial cells for zoomed-in areas at FOV center (d1)-(d2) and corner (d3)-(d4). 
}
\label{fig:dpc}
\end{figure}

As a first demonstration of snapshot multi-modal acquisition, we configured the M-FAST setup to acquire dual-channel fluorescence video over thin samples, since thick sample causes the ambiguity in the lateral shift. As sketched in Fig.~\ref{fig:fluourscent}(a), we inserted alternating red fluorescent emission filter (ET610/75m(Chroma), 12 mm) and green fluorescent emission filters (ET510/20m(Chroma), 12 mm) over successive micro-camera columns, allowing the M-FAST to simultaneously capture both spectral channels across the full system FOV (except at the left and right edges). As a proof-of-concept experiment, we imaged a mixture of \SI{10}{\micro\meter} red (590nm/605nm) and \SI{6}{\micro\meter} yellow-green microspheres (441nm/486nm). For fluorescence excitation, we used a high power \SI{470}{\nano\meter} blue LED (Chanzon, \SI{100}{\watt} Blue) with a short-pass excitation filter (Thorlabs, FES500). We stitched the two fluorescence channels separately to produce the composites in Fig.~\ref{fig:fluourscent}(b1-b2). M-FAST can clearly resolve the microsphere mixture, both spectrally and spatially. 

As presented above, two-channel imaging is achieved by inserting different filters in neighboring cameras with an overlapping FOV. More channels can be designed by increasing the amount of overlap between neighboring cameras (Fig.~\ref{fig:fluourscent}(a)). For four fluorescent channels, there are two options: (1) use rectangular image sensors with 75$\%$ FOV overlap along one dimension, or (2) using square sensors with 50$\%$ FOV overlap along both dimensions across the object plane. In general, the amount of FOV overlap, and thus the number of multimodal channels $c$ that an M-FAST system simultaneously captures, is a function of four key parameters: the sensor size ($s$), magnification of the micro-camera lens ($M$), desired number of cameras ($n$) and the total system FOV($F$). Within a simplified 1D analysis, the number of cameras along $x$ is given by $n_x = \frac{F_xM}{s_x}c_x-1$. The required camera array pitch can be shown to be given by $\frac{p_i}{s_i} = \frac{1}{Mc_i} > 1, i = x,y$. We assume the resolution is limited by the camera pixel size $\delta_x$. When giving a specific SBP image and a desired number of fluorescence channels, we can get required number of cameras using $n_x = \frac{SBP_x}{SBP_{cam,x}}c_x - 1$ by substituting $SBP_x = \frac{FM}{\delta_x}, SBP_{cam,x} = \frac{s_x}{\delta_x}$. Extension to a 2D camera array is achieved by multiplying the 1D expressions -- the total number of cameras is thus $n=n_xn_y$. 
Assuming square image sensors and reasonable M-FAST parameters ($SBP_{cam}$ = 10 MP), we plot $n$ against $c$ in Fig.~\ref{fig:fluourscent}(c). The blue line and orange line shows the relationship assuming 200-MP and 400-MP composite FOVs.

As a final demonstration of the flexibility of our novel multiscale microscope, we modified the illumination unit to include the ability to perform differential phase contrast (DPC) imaging~\cite{tian2015quantitative,xu2022tensorial}. Here,a 16-mm focal length condenser (ACL25416U, Thorlabs) and two subsequent condensers (ACL7560U, Thorlabs) are used for magnifying the active area of the LED source, shown in Fig.~\ref{fig:dpc}(a). We inserted an absorption mask at the illumination source conjugate plane to produce DPC modulation\cite{mehta2009quantitative}. In this design, the specimen is located at the Fourier plane of the DPC mask \cite{bonati2020phase}. To first quantitatively evaluate M-FAST DPC performance, we imaged a large collection of 10-\textmu m diameter polystyrene microspheres (refractive index: $n=1.59$) immersed in $n=1.58$ oil. To provide DPC modulation, we inserted four unique $0.8$ NA half-circle masks oriented along four cardinal directions and captured four snapshots. The resulting phase contrast maps are in Fig.~\ref{fig:dpc}(b), with average 1D traces through ten randomly selected beads shown in Fig.~\ref{fig:dpc}(c), both for beads at the micro-camera FOV center and edge. As reference we analytically derived the phase contrast profile of the microsphere, captured by a simplified single-objective DPC imaging with matching specifications.
The direct comparison between our experimental results and this analytical reference highlights the accurate performance of our system in this DPC mode. Next, we tested the effectiveness of M-FAST DPC imaging of buccal epithelial cells samples, which are mostly transparent and thus exhibit minimal structural contrast under a standard brightfield microscope. For this task, we employed ring-shaped DPC masks with 0.7 outer NA and 0.5 inner NA. 
Fig.~\ref{fig:dpc}(d) shows the zoom-ins of the associated DPC maps, which reveal detailed phase-specific features within cells. 

In summary, we have presented M-FAST, a new multi-modal, scalable imaging technique featuring a novel cascaded microscope design based on a finite-conjugate objective, a camera array and a fiber faceplate. This system provides fast acquisition of large-FOV images without scanning parts and without image vignetting. Moreover, our current configuration is versatile for different imaging modalities and features two fluorescence channels and a DPC module. We demonstrated that the system is scalable to increase the number of simultaneously observable fluorescence channels by tuning the camera array geometry. Our technique thus opens the door to high-speed, high-resolution, large-FOV, multi-modal microscopy.

\section{Funding} We acknowledge funding from NIH ORIP and NIH NIEHS for SBIR R44-OD24879, NIH NCI for SBIR 1R44CA250877, NIH NIBIB for SBIR 1R43EB030979


\section{Disclosures} MF: Ramona Optics Inc. (F,I,P,S), RH: Ramona Optics Inc. (F,I,P,S), KCZ: Ramona Optics Inc. (C).
\smallskip





\bibliographystyle{IEEEtran}
\bibliography{references}

\begin{thebibliography}{10}
\providecommand{\url}[1]{#1}
\csname url@samestyle\endcsname
\providecommand{\newblock}{\relax}
\providecommand{\bibinfo}[2]{#2}
\providecommand{\BIBentrySTDinterwordspacing}{\spaceskip=0pt\relax}
\providecommand{\BIBentryALTinterwordstretchfactor}{4}
\providecommand{\BIBentryALTinterwordspacing}{\spaceskip=\fontdimen2\font plus
\BIBentryALTinterwordstretchfactor\fontdimen3\font minus
  \fontdimen4\font\relax}
\providecommand{\BIBforeignlanguage}[2]{{%
\expandafter\ifx\csname l@#1\endcsname\relax
\typeout{** WARNING: IEEEtran.bst: No hyphenation pattern has been}%
\typeout{** loaded for the language `#1'. Using the pattern for}%
\typeout{** the default language instead.}%
\else
\language=\csname l@#1\endcsname
\fi
#2}}
\providecommand{\BIBdecl}{\relax}
\BIBdecl

\bibitem{sdobnov2018recent}
A.~Y. Sdobnov, M.~Darvin, E.~Genina, A.~Bashkatov, J.~Lademann, and V.~Tuchin,
  ``Recent progress in tissue optical clearing for spectroscopic application,''
  \emph{Spectrochimica Acta Part A: Molecular and Biomolecular Spectroscopy},
  vol. 197, pp. 216--229, 2018.

\bibitem{kim2016simultaneous}
C.~K. Kim, S.~J. Yang, N.~Pichamoorthy, N.~P. Young, I.~Kauvar, J.~H. Jennings,
  T.~N. Lerner, A.~Berndt, S.~Y. Lee, C.~Ramakrishnan \emph{et~al.},
  ``Simultaneous fast measurement of circuit dynamics at multiple sites across
  the mammalian brain,'' \emph{Nature methods}, vol.~13, no.~4, pp. 325--328,
  2016.

\bibitem{fan2019video}
J.~Fan, J.~Suo, J.~Wu, H.~Xie, Y.~Shen, F.~Chen, G.~Wang, L.~Cao, G.~Jin, Q.~He
  \emph{et~al.}, ``Video-rate imaging of biological dynamics at centimetre
  scale and micrometre resolution,'' \emph{Nature Photonics}, vol.~13, no.~11,
  pp. 809--816, 2019.

\bibitem{thomson2021gigapixel}
E.~Thomson, M.~Harfouche, P.~Konda, C.~W. Seitz, K.~Kim, C.~Cooke, S.~Xu,
  R.~Blazing, Y.~Chen, W.~S. Jacobs \emph{et~al.}, ``Gigapixel behavioral and
  neural activity imaging with a novel multi-camera array microscope,''
  \emph{bioRxiv}, 2021.

\bibitem{krishnamurthy2019scale}
D.~Krishnamurthy, H.~Li, F.~B. du~Rey, P.~Cambournac, A.~Larson, and
  M.~Prakash, ``Scale-free vertical tracking microscopy: Towards bridging
  scales in biological oceanography,'' \emph{bioRxiv}, p. 610246, 2019.

\bibitem{JOHNSON202070}
``Probabilistic models of larval zebrafish behavior reveal structure on many
  scales,'' \emph{Current Biology}, vol.~30, no.~1, pp. 70--82.e4, 2020.

\bibitem{nguyen2016whole}
J.~P. Nguyen, F.~B. Shipley, A.~N. Linder, G.~S. Plummer, M.~Liu, S.~U. Setru,
  J.~W. Shaevitz, and A.~M. Leifer, ``Whole-brain calcium imaging with cellular
  resolution in freely behaving caenorhabditis elegans,'' \emph{Proceedings of
  the National Academy of Sciences}, vol. 113, no.~8, pp. E1074--E1081, 2016.

\bibitem{park2021review}
J.~Park, D.~J. Brady, G.~Zheng, L.~Tian, and L.~Gao, ``Review of bio-optical
  imaging systems with a high space-bandwidth product,'' \emph{Advanced
  Photonics}, vol.~3, no.~4, p. 044001, 2021.

\bibitem{konda2020fourier}
P.~C. Konda, L.~Loetgering, K.~C. Zhou, S.~Xu, A.~R. Harvey, and R.~Horstmeyer,
  ``Fourier ptychography: current applications and future promises,''
  \emph{Optics express}, vol.~28, no.~7, pp. 9603--9630, 2020.

\bibitem{dai2022quantitative}
X.~Dai, S.~Xu, X.~Yang, K.~C. Zhou, C.~Glass, P.~C. Konda, and R.~Horstmeyer,
  ``Quantitative jones matrix imaging using vectorial fourier ptychography,''
  \emph{Biomedical optics express}, vol.~13, no.~3, pp. 1457--1470, 2022.

\bibitem{xu2021imaging}
S.~Xu, X.~Dai, X.~Yang, K.~Zhou, P.~Konda, and R.~Horstmeyer, ``Imaging
  anisotropy with vectorial fourier ptychography,'' in \emph{Computational
  Optics 2021}, vol. 11875.\hskip 1em plus 0.5em minus 0.4em\relax SPIE, 2021,
  p. 118750F.

\bibitem{voigt2019mesospim}
F.~F. Voigt, D.~Kirschenbaum, E.~Platonova, R.~A. Campbell, R.~Kastli,
  M.~Schaettin, L.~Egolf, A.~Van~der Bourg, P.~Bethge, K.~Haenraets
  \emph{et~al.}, ``The mesospim initiative: open-source light-sheet microscopes
  for imaging cleared tissue,'' \emph{Nature methods}, vol.~16, no.~11, pp.
  1105--1108, 2019.

\bibitem{he2020snapshot}
K.~He, X.~Wang, Z.~W. Wang, H.~Yi, N.~F. Scherer, A.~K. Katsaggelos, and
  O.~Cossairt, ``Snapshot multifocal light field microscopy,'' \emph{Optics
  Express}, vol.~28, no.~8, pp. 12\,108--12\,120, 2020.

\bibitem{mcconnell2016novel}
G.~McConnell, J.~Tr{\"a}g{\aa}rdh, R.~Amor, J.~Dempster, E.~Reid, and W.~B.
  Amos, ``A novel optical microscope for imaging large embryos and tissue
  volumes with sub-cellular resolution throughout,'' \emph{Elife}, vol.~5, p.
  e18659, 2016.

\bibitem{brady2012multiscale}
D.~J. Brady, M.~E. Gehm, R.~A. Stack, D.~L. Marks, D.~S. Kittle, D.~R. Golish,
  E.~Vera, and S.~D. Feller, ``Multiscale gigapixel photography,''
  \emph{Nature}, vol. 486, no. 7403, pp. 386--389, 2012.

\bibitem{zheng20140}
G.~Zheng, X.~Ou, and C.~Yang, ``0.5 gigapixel microscopy using a flatbed
  scanner,'' \emph{Biomedical optics express}, vol.~5, no.~1, pp. 1--8, 2014.

\bibitem{liu2021meta}
Y.~Liu, Q.-Y. Yu, Z.-M. Chen, H.-Y. Qiu, R.~Chen, S.-J. Jiang, X.-T. He, F.-L.
  Zhao, and J.-W. Dong, ``Meta-objective with sub-micrometer resolution for
  microendoscopes,'' \emph{Photonics Research}, vol.~9, no.~2, pp. 106--115,
  2021.

\bibitem{stamenov2014panoramic}
I.~Stamenov, A.~Arianpour, S.~J. Olivas, I.~P. Agurok, A.~R. Johnson, R.~A.
  Stack, R.~L. Morrison, and J.~E. Ford, ``Panoramic monocentric imaging using
  fiber-coupled focal planes,'' \emph{Optics express}, vol.~22, no.~26, pp.
  31\,708--31\,721, 2014.

\bibitem{boas2010laser}
D.~A. Boas and A.~K. Dunn, ``Laser speckle contrast imaging in biomedical
  optics,'' \emph{Journal of biomedical optics}, vol.~15, no.~1, p. 011109,
  2010.

\bibitem{xu2021imaging2}
S.~Xu, X.~Yang, W.~Liu, J.~Jonsson, R.~Qian, P.~C. Konda, K.~C. Zhou, Q.~Dai,
  H.~Wang, E.~Berrocal \emph{et~al.}, ``Imaging dynamics beneath turbid media
  via parallelized single-photon detection,'' \emph{arXiv preprint
  arXiv:2107.01422}, 2021.

\bibitem{xu2022transient}
S.~Xu, W.~Liu, X.~Yang, J.~J{\"o}nsson, R.~Qian, P.~McKee, K.~Kim, P.~C. Konda,
  K.~C. Zhou, L.~Krei{\ss} \emph{et~al.}, ``Transient motion classification
  through turbid volumes via parallelized single-photon detection and deep
  contrastive embedding,'' \emph{Frontiers in neuroscience}, vol.~16, 2022.

\bibitem{PTGui}
\BIBentryALTinterwordspacing
{New House Internet Services BV}, ``Ptgui.'' [Online]. Available:
  \url{https://ptgui.com/}
\BIBentrySTDinterwordspacing

\bibitem{fujimaki2014reduction}
Y.~Fujimaki and H.~Taniguchi, ``Reduction of speckle contrast in multimode
  fibers using piezoelectric vibrator,'' in \emph{Laser Resonators,
  Microresonators, and Beam Control XVI}, vol. 8960.\hskip 1em plus 0.5em minus
  0.4em\relax International Society for Optics and Photonics, 2014, p. 89601S.

\bibitem{huang2020retrieving}
G.~Huang, D.~Wu, J.~Luo, Y.~Huang, and Y.~Shen, ``Retrieving the optical
  transmission matrix of a multimode fiber using the extended kalman filter,''
  \emph{Optics express}, vol.~28, no.~7, pp. 9487--9500, 2020.

\bibitem{tian2015quantitative}
L.~Tian and L.~Waller, ``Quantitative differential phase contrast imaging in an
  led array microscope,'' \emph{Optics express}, vol.~23, no.~9, pp.
  11\,394--11\,403, 2015.

\bibitem{xu2022tensorial}
S.~Xu, X.~Dai, X.~Yang, K.~C. Zhou, K.~Kim, V.~Pathak, C.~Glass, and
  R.~Horstmeyer, ``Tensorial tomographic differential phase-contrast
  microscopy,'' \emph{arXiv preprint arXiv:2204.11397}, 2022.

\bibitem{mehta2009quantitative}
S.~B. Mehta and C.~J. Sheppard, ``Quantitative phase-gradient imaging at high
  resolution with asymmetric illumination-based differential phase contrast,''
  \emph{Optics letters}, vol.~34, no.~13, pp. 1924--1926, 2009.

\bibitem{bonati2020phase}
C.~Bonati, T.~Laforest, M.~Kunzi, and C.~Moser, ``Phase sensitivity in
  differential phase contrast microscopy: limits and strategies to improve
  it,'' \emph{Optics Express}, vol.~28, no.~22, pp. 33\,767--33\,783, 2020.

\end{thebibliography}

\end{document}